\documentclass[conference,a4paper]{IEEEtran}

\usepackage{flushend}
\usepackage{setspace}
 \def\ps@headings{%
 \def\@oddhead{\mbox{}\scriptsize\rightmark \hfil \thepage}%
 \def\@evenhead{\scriptsize\thepage \hfil \leftmark\mbox{}}%
 \def\@oddfoot{}%
 \def\@evenfoot{}}
 \makeatother
\usepackage{amssymb}
\usepackage{amsmath}
\usepackage[dvips]{graphicx}
\usepackage{enumerate}

\IEEEoverridecommandlockouts

\usepackage{color}
\usepackage{cite}
\usepackage{verbatim}
\usepackage{url}

\def\backGroundColor{white}
\def\txtsize{\normalsize}
\def\bl{\par\textcolor{\backGroundColor}{\txtsize{ empty line }}\par}
\def\it{\textit}

\def\includeComments{include}
\def\includ{include}

\def\comm[#1]{\ifx\includeComments\includ  \bl \texttt{\textbf{\textit{Note: #1}}} \par \fi}
\def\inlinecomm[#1]{\ifx\includeComments\includ  \textit{Note: #1} \fi}

\def\w{\omega}
\def\ckp{C'_k}
\def\hii{h_{ii}}

\def\hkk{h_{kk}}
\def\hij{h_{ij}}
\def\hji{h_{ji}}
\def\hoo{h_{11}}
\def\htt{h_{22}}
\def\hot{h_{12}}
\def\hto{h_{21}}

\def\blackbox{\hfill {\vrule height6pt width6pt depth0pt}}
\def\Box{\hfill \framebox(5.25,5.25){}}

\def\QD{\blackbox}

\newcommand{\para}[1]{\medskip \noindent {\bf #1}}
\newcommand{\softpara}[1]{\smallskip \noindent \underline{#1}}

\newcommand{\blue}[1]{\textcolor{black}{#1}}

\newcommand{\magneta}[1]{}

\newcommand{\ommited}[1]{}

\newcommand{\cbl}{\color{black}}

\newcommand{\cb}{\color{black}}

\newcommand{\Eat}[1]{}
\newcommand{\eat}[1]{}

\newtheorem{corollary}{Corollary}

\newtheorem{defin}{Definition}
\newtheorem{ex}{EXAMPLE}
\newtheorem{thm}{Theorem}
\newtheorem{lem}{Lemma}
\newtheorem{alg}{Algorithm}
\newtheorem{ob}{Observation}

\newenvironment{cor-prf}{\begin{corollary} \nopagebreak}{\end{corollary}}
\newenvironment{theorem}{\begin{thm} \nopagebreak}{{\hfill$\blackbox$} \end{thm}}
\newenvironment{thm-prf}{\begin{thm} \nopagebreak}{\end{thm}}
\newenvironment{lemma}{\begin{lem} \nopagebreak}{{\hfill$\blackbox$} \end{lem}}
\newenvironment{lem-prf}{\begin{lem} \nopagebreak}{\end{lem}}

\Eat{

}

\newcommand{\squishlist}{
 \begin{list}{$\bullet$}
  { \setlength{\itemsep}{0pt}
     \setlength{\parsep}{3pt}
     \setlength{\topsep}{3pt}
     \setlength{\partopsep}{0pt}
     \setlength{\leftmargin}{1.5em}
     \setlength{\labelwidth}{1em}
     \setlength{\labelsep}{0.5em} } }

\newcommand{\squishlisttwo}{
 \begin{list}{$\bullet$}
  { \setlength{\itemsep}{0pt}
     \setlength{\parsep}{0pt}
    \setlength{\topsep}{0pt}
    \setlength{\partopsep}{0pt}
    \setlength{\leftmargin}{2em}
    \setlength{\labelwidth}{1.5em}
    \setlength{\labelsep}{0.5em} } }

\newcommand{\squishend}{
  \end{list}  }

\author{
\IEEEauthorblockN{Navid Hamedazimi, Himanshu Gupta}
    \vspace*{-3mm}
    Stony Brook University, NY
}

\begin{document}

\title{Optimal Spectrum Management in Two-User Interference Channels}
\maketitle
\begin{spacing}{0.9631}

\begin{abstract}
In this work, we address the problem of optimal spectrum management in
continuous frequency domain in multiuser interference channels. The
objective is to maximize the weighted sum of user capacities. Our main
results are as follows: (i) For frequency-selective channels, we prove
that in an optimal solution, each user uses maximum power; this result
also generalizes to the cases where the objective is to maximize the
weighted {\em product} (i.e., proportional fairness) of user
capacities.  (ii) For the special case of two users in flat channels,
we solve the problem optimally.

\end{abstract}

\section{\bf Introduction}
\label{sec:intro}

In this paper, we address the problem of maximizing weighted sum of user
capacities in multiuser communication systems in a common frequency
band. We consider a continuous frequency domain. For
frequency-selective channels, we prove that in an optimal solution,
each user must use the maximum power available to it. This
maximum-power result also holds in the case wherein the objective is
to maximize the weighted product of user capacities; this objective is
generally used to achieve proportional fairness. For the special case
of two users in flat channels, we present an optimal spectrum
management solution.

In a multiuser communication system~\cite{msn,matrix,join}, users either have to partition
the available frequency ({\em FDMA}), or use {\em frequency sharing}
(i.e., each user uses the entire spectrum), or a combination of the
two (i.e., use partially-overlapping spectrums).  Intuitively, FDMA is
the optimal answer in the case of strong cross coupling (also referred
to as strong interference scenario), and frequency sharing is optimal
when the cross coupling is very weak. In the intermediate case, the
optimal solution may be a combination of the two
strategies~\cite{UCLApaper} (i.e., users may use partially-overlapping
spectrums).

There exist an extensive literature on the effect of cross coupling on
choosing between FDMA and frequency sharing. The works in~\cite{ucb}
and~\cite{umn-2} provide sufficient conditions under which FDMA is
guaranteed to be optimal; these conditions are group-wise conditions,
i.e., each pair of users need to satisfy the condition.
Recently, Zhao and Pottie~\cite{UCLApaper} derived a tight condition
which when satisfied by a pair of users guarantees that the given pair
uses orthogonal frequencies (i.e, FDMA for the pair). Their result
holds for any pareto optimal solution.

In the general interference scenarios in multiuser systems, the
weighted sum-rate maximization problem is a non-convex optimization
problem, and is generally hard to solve~\cite{yu2002distributed}.
However, two general approaches have been proposed: (i) One approach
considers the Lagrangian dual problem decomposed in frequency after
first descretizing the spectrum~\cite{yu2006dual}; the resulting
Lagrangian dual problem is convex and potentially easier to
solve~\cite{cendrillon2006optimal,umn-1}. More
importantly,~\cite{umn-1} proves that the duality gap goes to zero
when the number of ``sub-channels'' goes to infinity. However, the
time-complexity of their method is a high-degree polynomial in the
number of sub-channels (thus, becoming prohibitively expensive for the
continuous frequency domain problem).  (ii) The second approach
changes the formulation of the problem to get an equivalent primal
domain convex maximization problem~\cite{UCLApaper}. Eventhough, the
above approaches almost reduce the spectrum management problem to a
convex optimization problem, they fall short of designing an optimal
or approximation algorithm with bounded convergence.

\blue{The recent works in~\cite{UCLApaper,otherUCLA} find the optimal
  solution for the special case of two ``symmetric'' users; their
  result is very specific, and doesn't generalize to weighted or
  non-symmetric links. In another insightful work,~\cite{hong} gives a
  characterization of the optimal solution for the two-user case which
  essentially yields a four to six variable equation. Our work
  essentially improves on these results and solves the problem for the
  general case of two users, using an entirely different technique.}

\softpara{Discrete Frequency Spectrum Management.}  In other related
works,~\cite{umn-1} and~\cite{umn-11} consider the spectrum management
problem in {\em discrete} frequency domain, wherein the available
spectrum is already divided into {\em given} orthogonal channels and
user power spectral densities are constant in each channel. Their
motivation for considering the discrete version is to facilitate a
numerical solution~\cite{umn-1}.
The discrete version is shown to be NP-hard (even for two users), and
in~\cite{umn-2} the authors give a {\em sufficient} condition for the
optimal to be an FDMA solution. Even when restricted to FDMA
solutions, they observe that the discrete version remains
inapproximable, but provide a PTAS~\cite{umn-11} for the continuous
version (when restricted to FDMA solutions).
Note that, for two users, the discrete version remains
NP-hard~\cite{umn-2}, while the continuous version has been solved
optimally in our paper (Section~\ref{sec:two}). Thus, discretizing the
spectrum seems to make the spectrum allocation problem only harder,
contrary to the motivation in~\cite{umn-1}.  Moreover, discretization
of a given spectrum can actually reduce achievable capacity.

\para{Our Results.}  In this paper, we address the following spectrum
management problem: Given a spectrum band of width $W$ and a set of
$n$ users each with a maximum transmit power, the SAPD (spectrum
allocation and power distribution) problem is to determine power
spectrum densities of the users in the continuous frequency domain to
maximize the weighted sum of user capacities (as computed by the
generalized Shannon-Hartley theorem). For the above SAPD problem, we
present the following results.

\begin{itemize}
\item
For frequency-selective channels, we show that in an optimal SAPD
solution, each user must use the maximum transmit power. We extend the
result to the cases wherein the objective is to maximize the weighted
product of user capacities.

\item
For the special case of two users in flat channels, we design an
optimal solution for the SAPD problem. This is a direct improvement of
the recent recent in~\cite{UCLApaper} which solves the problem
optimally for the special case of two users with symmetric (equal
channel gains and noise) and flat channels.
\end{itemize}

\section{\bf Problem Formulation, and Notations}
\label{sec:model}

\para{Model, Terms, and Notations.}  We are given a set of users $i$
(formed by a transmitter $s_i$ and a receiver $r_i$) and a frequency
spectrum $[0,W]$. \eat{Without loss of generality, we assume that the
users do not share any nodes.} The background noise at the {\em
receiver} of user $i$ is assumed to be white, i.e., constant across
the spectrum, and has a constant value of $N_i$ (Watts/Hz) at each
frequency.  We use $\hij(x)$ to denote {\em channel gain} between the
{\em sender} of user $i$ and the {\em receiver} of user $j$ at frequency $x$.
\eat{in practice, $\hij$ can be determined by the distance between the
sender $s_i$ and receiver $r_j$.}

\softpara{Power Spectrum Density (PSD) $p_i(x)$; Total Power.} For a
user $i$, the {\em power spectral density (PSD)} is a function $p_i:
[0,W] \mapsto \mathbb{R}_{\leq0}$ that gives the power at each
frequency of the signal used by the transmitter $s_i$ to communicate
with its receiver $r_i$. Thus, $p_i(x)$ is the power of $s_i$'s signal
at frequency $x$. In this paper, we allow arbitrary PSD functions. The
{\em total power} used by a user $i$ is given by $\int\limits_0^W
p_i(x) dx.$

\softpara{Maximum Total Power.} Each user $i$ is associated
with a {\em maximum total power} $P_i$, which is the bound on the
total power used by its transmitter $s_i$. That is, each PSD function
$p_i(x)$ must satisfy the below condition:
\begin{equation}
\int\limits_0^W p_i(x) dx \leq P_i. \label{eqn:P}
\end{equation}

\softpara{Spectrum Used.} Given a PSD function $p_i(x)$ for a user
$i$, the {\em spectrum used} by user $i$ is defined as $\{x | p_i(x) >
0\}$, i.e., the set of frequencies wherein the power is non-zero.
Thus, {\em {\bf disjoint}} spectrums are orthogonal.

\softpara{User Capacity.} Given PSD functions $\{p_i(x)\}$ for a set
of users in a communication system, the (maximum achievable rate) capacity
$C_i$ of a user $i$ can be determined using the generalized
Shannon-Hartly theorem as below. Here, we assume that the signals to
be Gaussian processes, and treat interference as noise, as in prior
works~\cite{umn-1,umn-11,umn-2,ucb}.
\begin{equation}
C_i = \int\limits_0^W {\log \left(1+\frac{p_i(x)\hii(x)}{I_i(x) + N_i} \right)dx}. \label{eqn:sh}
\end{equation}
Above, $\hii$ is
the channel gain, and $I_i(x)$ is the total interference on
frequency $x$ at the receiver $r_i$ due to other users. The
interference $I_i(x)$ is computed as follows.
$$I_i(x) = \sum_{j \neq i} p_j(x)\hji(x).$$

\para{Spectrum Allocation and Power Distribution (SAPD) Problem.}
Given a set of users $\{1,2,\ldots,n\}$, maximum total power values
$P_i$ for each user $i$, noise $N_i$ at each receiver $r_i$, and an
available frequency spectrum $[0,W]$, the {\em Spectrum Allocation and
  Power Distribution (SAPD)} problem is to determine the PSD functions
$\{p_i(x)\}$ for the given users such that the {\em total (system)
  weighted capacity} $\sum_i \w_i C_i$ is maximized where $\w_i$ are
the given weights, under the constraint of Equation~\ref{eqn:P} (i.e.,
the total power used by each user $i$ is at most $P_i$). Note that
determination of PSD functions also gives the allocation of spectrum
across users (i.e., spectrums used by each user).


\section{\bf Optimal SAPD Solution Uses Maximum Power}
\label{sec:power}

In this section, we prove that in an optimal SAPD solution, each user
uses maximum total power. We note that our result does {\em not}
contradict the prior ``binary-power control'' 
results of~\cite{stanfordPower,otherPower,binary-ggok,binary-emk} 
who consider a different and restricted model. In 
particular, they consider a model wherein each user
uses a constant PSD across the available spectrum (i.e., each user
either uses the {\em entire} spectrum with a constant PSD or remains
silent). For this model, they show that to achieve maximum sum of user
rates either (i) each user uses maximum power, or (ii) one of the
users is silent (with the other user using maximum power). In
contrast, in our model (wherein each user can use an arbitrary PSD
function, and thus, an arbitrary subset of the spectrum), we show that
each user must use maximum power to achieve maximum sum of user
capacities. In fact, it is easy to see from our
Lemma~\ref{lem:two-disjoint} that, in our model, the sum of rates
achieved when one user is silent is {\em always} sub-optimal.

\begin{thm-prf}
\label{thm:multi-power}
For frequency-selective channels, in an optimal SAPD solution, each
user uses maximum power, i.e., for each user $i$, $\int\limits_0^W
p_i(x) dx = P_i$.
\end{thm-prf}

\begin{proof}
Let $n$ be the number of users. Consider an optimal solution
$\{p_i(x)\}$, where $p_i(x)$ is the PSD of the $i^{th}$ user. Assume
that the claim of the theorem doesn't hold, i.e., there is a user $k$
such that
$$p' = P_k - \int_{0}^{W}{p_k(x)dx} > 0.$$
\noindent
Below, we use $p'$ to improve on the given solution, which will contradict
our assumption that the given solution is optimal and thus, proving the
theorem.

Now, for an appropriate constant $\epsilon$ (as determined later), we
change the given optimal solution as follows.
\begin{itemize}
\item
First, in the spectrum $[0,\epsilon]$, we power-off all the users,
i.e., for all $i$, we set $p_i(x)=0$ for $x \in [0,\epsilon]$.

\item
Second, we uniformly add the power $p'$ to $k$'s PSD in the spectrum
$[0,\epsilon]$, i.e., we set $p_k(x)$ to $p'/\epsilon$ for $x \in [0,
  \epsilon]$.
\end{itemize}
The first change causes a decrease in the capacity of every user
(including $k$), while the second change results in some new capacity
for $k$. We can compute these amounts as follows.
\begin{itemize}
\item
The decrease $\bigtriangledown_i$ in capacity of each user $i$ (including
$k$) due to the changes can be computed as:
\begin{align}
\bigtriangledown_i &=& \int_0^{\epsilon} \log \left(1+ \frac{p_i(x)\hii(x)}{I_i(x) + N_i} \right) dx \nonumber \\
	&\leq& \epsilon \log (1+ \frac{p_{max}h_{max}}{N_{min}})  \label{eqn:dec}
\end{align}
Above, $N_{min} = \min_i N_i$, $p_{max}= \max_{i,x} p_i(x)$, and
$h_{max} = \max_{i,x} \hii(x) $, where $x \in [0,\epsilon]$ and $i$
varies over all users.

\item
The new capacity $\ckp$ of user $k$ in $[0,\epsilon]$ after the second
change is:
\begin{eqnarray}
\ckp &=& \int_0^{\epsilon} \log \left(1+ \frac{(p'/\epsilon)\hkk(x)}{N_k} \right) dx \nonumber \\
&\geq&\epsilon \log(1+ \frac{p'h_{min}}{N_k\epsilon}) \label{eqn:inc}
\end{eqnarray}
Above, we have \blue{used} $h_{min} = \min_{x} \hkk(x)$. 
\end{itemize}
Now, the overall increase in the sum of weighted capacities of all the users
is $$\w_k\ckp - \sum_i \w_i \bigtriangledown_i.$$ Below, we pick an $\epsilon$
that will ascertain $\w_k\ckp > \w_i\sum_i \bigtriangledown_i$. Such an
$\epsilon$ will imply that the above suggested changes result in an
increase in the weighted sum of user capacities, and thus, proving the
theorem. In particular, using Equation~\ref{eqn:dec}
and~\ref{eqn:inc}, we pick an $\epsilon$ such that:
\begin{eqnarray*}
\w_k \epsilon \log(1+ \frac{p'h_{min}}{N_k\epsilon}) & > & \epsilon \sum_{i} \w_i \log (1+ \frac{p_{max}h_{max}}{N_{min}}) \\
\log(1+ \frac{p'h_{min}}{N_k\epsilon}) & > & (\frac{\sum_i \w_i}{\w_k}) \log (1+ \frac{p_{max}h_{max}}{N_{min}}) \\
1+ \frac{p'h_{min}}{N_k\epsilon}  & > & (1+ \frac{p_{max}h_{max}}{N_{min}})^{(\frac{\sum_i \w_i}{\w_k})}\\
\epsilon & < & \frac{p'h_{min}}{N_k((1+\frac{p_{max}h_{max}}{N_{min}})^{(\frac{\sum_i \w_i}{\w_k})}-1)}.
\end{eqnarray*}
Since the above expression is positive, there exists an $\epsilon$ for
which the above suggested changes result in an increase in the weighted
sum of user capacities. This contradicts the assumption that the original
solution is optimal, and thus, proving the theorem.
\end{proof}

Theorem~\ref{thm:multi-power} can be easily generalized to the case
wherein the objective is to maximize the weighted {\em product} of user
capacities, i.e., to achieve proportional fairness. We defer the proof
to Appendix~\ref{app:fair}.

\begin{theorem}
\label{thm:multi-power-fair}
For the SAPD problem wherein the objective is to maximize the weighted
product of user capacities, the optimal solution uses maximum power
for each user.
\end{theorem}

\eat{
\begin{cor-prf}
\label{cor:weighted-power}
For the {\em weighted} SAPD problem wherein the objective is to maximize the 
{\em weighted} sum of user capacities, the optimal solution uses maximum power 
for each user.
\end{cor-prf}
\begin{proof}
For the weighted sum case, let the objective be to maximize $\sum_i
w_i C_i$, where $w_i \geq 0$ are the given weights. Following the
proof of the above theorem, here, we need to pick an $\epsilon$ that
will ascertain $w_k\ckp > \sum_i w_i\bigtriangledown_i$.  Thus,
$$\epsilon  <  \frac{p'h_{min}}{N_k((1+\frac{p_{max}h_{max}}{N_{min}})^{w}-1)}$$
sufficies where $w = (\sum_i w_i)/w_k$.
\end{proof}

See Appendix~\ref{app:fair} for the proof.
}

\section{\bf Optimal SAPD Solution for Two Users in Flat Channels}
\label{sec:two}

In this section, we present an optimal solution for the SAPD problem
for the special case of two users in flat channels.  We use $h_{ij}$
to denote the channel gain, i.e., $h_{ij}(x) = h_{ij}$ for all
$x$. {\em For clarity of presentation, in this section, we implicitly
  assume the given weights $\w_i$ to be uniform and unit; the
  generalization of our results to non-uniform weights is
  straightforward.}

We start with an important lemma. The lemma's proof is very tedious
(see Appendix~\ref{app:rect}).

\begin{lemma}
\label{lem:rect}
For a two user SAPD problem in flat channels, there exists an optimal
solution wherein the PSD of each user is constant in the spectrum
shared by the users. More formally, there exists an optimal solution
such that if $S_1$ and $S_2$ are the spectrums used by the respective
users, then for $x \in (S_1 \cap\ S_2)$, $p_i(x) = c_i$ for some
constants $c_i$ ($i$ = 1,2).
\end{lemma}

A somewhat related result from~\cite{ucb} states that any SAPD
solution for $n$ users can be expressed using piecewise-constant PSD's
over appropriate 2$n$ pieces of the available spectrum; this result
requires 4 pieces for $n=2$ users. In contrast, our above lemma
implies a stronger result for an SAPD solution for two users, and is
essential to our \blue{result.}

\para{Optimal SAPD Solution for Two Users.}  Consider a system with
two users and an available spectrum $[0.W]$. The optimal SAPD solution
can take three possible forms, viz., (i) the users use disjoint
subspectrums, (ii) both users use the same subspectrum, (iii) the
users use {\em partially-overlapping} (i.e., non-disjoint and
non-equal) subspectrums. We can solve the first and the second cases
optimally by using the below Lemmas~\ref{lem:two-disjoint}
and~\ref{lem:full-overlap} respectively. We defer the proofs
to Appendix~\ref{app:disjoint}, but Lemma~\ref{lem:two-disjoint} is a slight
generalization of a result from~\cite{gummadi-hotnets-08} while
Lemma~\ref{lem:full-overlap} follows easily from Equation~\ref{eqn:sh}
and Lemma~\ref{lem:rect}.

\begin{lemma}
\label{lem:two-disjoint}
Consider a system of two users $\{1,2\}$, and an available spectrum
$[0,W]$. If the spectrums used by the two users are disjoint, then the
maximum system capacity is
$$W \log (1+\frac{P_1\hoo}{WN_1} +\frac{P_2\htt}{WN_2}),$$
and is achieved by dividing the spectrum in the ratio $N_{2}P_{1}\hoo
: N_{1}P_{2}\htt$.
\end{lemma}

It is easy to see from the above lemma that the system
capacity obtained when one of the users is silent is always
less than that obtained by the partitioning the spectrum as
suggested in the lemma.

\begin{lemma}
\label{lem:full-overlap}
Consider a system with two users, and an available spectrum
$[0,W]$. If the spectrums used by the two users is equal, then the maximum
system capacity possible is:
$$W \log (1+\frac{P_1\hoo}{P_2\hto +WN_1} ) +  W \log (1+\frac{P_2\htt}{P_1\hot +WN_2 } ).$$
\end{lemma}

In the following paragraph, we show how to compute an
optimal solution for the remaining third case, viz., wherein
users use partially-overlapping subspectrums. The overall optimal
SAPD solution can be then computed by taking the best
of the optimal solutions for the above three cases.

\begin{figure}
\hspace{0.2in}
\includegraphics[width=4.5in]{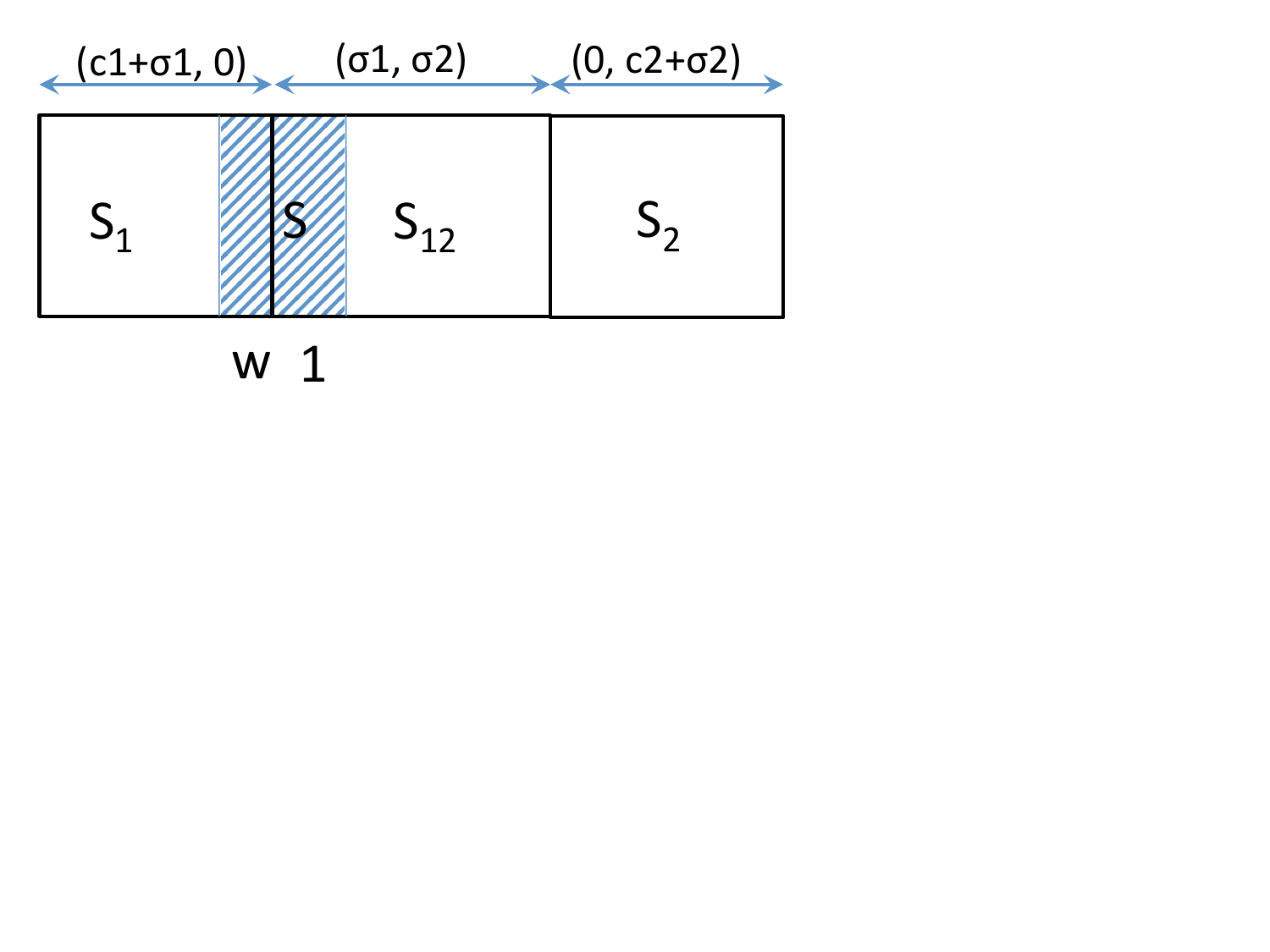}
\vspace*{-2.2in}
\caption{$S_1$ and $S_2$ are subspectrums used exclusively by users 1 and 2
respectively, and $S_{12}$ is the subspectrum used by both the users. The shaded
part of the spectrum is $S$ (used to derive the final two equations) and is
composed of two subspectrums of width 1 and $w$ respectively.
The top of the figures denotes the PSDs used by the users, e.g.,
$(c1+\sigma_1,0)$ signifies that the PSD values of the two users is
$c1+\sigma_1$ and $0$ respectively in $S_1$.}
\vspace*{-0.2in}
\label{fig:overlap}
\end{figure}

\para{Optimal Partially-Overlapping SAPD Solution.}  Consider an SAPD
solution that is optimal among all partially-overlapping SAPD
solutions. In such a solution, the available spectrum can be divided
into three subspectrums $S_1$, $S_2$, and $S_{12}$, where $S_1$ and
$S_2$ are used exclusively by user 1 and 2 respectively and $S_{12}$
is used by both the users. \blue{We assume $S_1$ and $S_2$ to be
  non-zero; the cases wherein one of them is zero are easier (see Appendix~\ref{app:cases}.}
Now, since the noise is white, we can assume without loss of
generality, that these three subspectrums are contiguous.
It is easy to see that each user 1 must use a constant PSD in $S_1$,
and user 2 must use a constant PSD in $S_2$. Also, by
Lemma~\ref{lem:rect}, we know that each user must use a constant
PSD in $S_{12}$, and each of the three subspectrums. Finally, by
Lemma~\ref{lem:diff-inter} (see Appendix~\ref{app:disjoint}), the PSD of user 1 in $S_1$
must be greater than its PSD in $S_{12}$; similarly, the PSD of user 2
in $S_2$ must be greater than its PSD in $S_{12}$. Now, let $\sigma_1$
and $\sigma_2$ be the PSD's in $S_{12}$ of user $1$ and $2$
respectively, $\sigma_1+c_1$ be the PSD of user 1 in $S_1$, and
$\sigma_2+c_2$ be the PSD of user $2$ in $S_2$. See
Figure~\ref{fig:overlap}. The total system capacity can now
be written as follows.
\[
\begin{array}{ll}
    B\ =\ &S_1\log(1+\frac{(\sigma_1+c1)\hoo}{N_1})+S_2\log(1+\frac{(\sigma_2+c_2)\htt}{N_2})+\\
      &S_{12}(\log(1+\frac{\sigma_1\hoo}{\sigma_2+N_1})+\log(1+\frac{\sigma_2\htt}{\sigma_1+N_2}))
\end{array}
\]
To find the optimal SAPD solution of the above form, we need to
essentially find values of the seven variables
$S_1,S_2,S_{12},\sigma_1,c_1,\sigma_2$ and $c_2$ such that the above
$B$ is maximized. We do so by determining six independent equations
that must hold true for an optimal $B$.  These six equations will help
us eliminate all but one of the seven variables in $B$, yielding a
formulation of $B$ in terms of a single variable. We can then
differentiate $B$ with respect to the remaining variable, find the
root of the differential equation equated to zero, and thus, determine
the value of all the seven variables. Below, we derive the six
equations (Equations~\ref{eqn:first} to~\ref{eqn:sixth}) that relate
the above seven variables. Below, $S_1$, $S_2$ and $S_{12}$ refer to the
{\em sizes} of the corresponding spectrums.

\begin{itemize}

\item
Since $W$ is the size of the total available spectrum, we have (by a
simple application of Lemma~\ref{lem:two-disjoint}):
\begin{equation}
W = S_1+S_2+S_{12} \label{eqn:first}
\end{equation}

\item
Since $P_1$ and $P_2$ are the maximum total power of users $1$ and $2$ respectively, by Theorem~\ref{thm:multi-power},
we have:
\begin{eqnarray}
P_1 &=& S_1(\sigma_1+c_1)+S_{12}\sigma_1 \label{eqn:second} \\
P_2 &=& S_2(\sigma_2+c_2)+S_{12}\sigma_2 \label{eqn:third}
\end{eqnarray}

\item
Note that the PSD's of the users 1 and 2 in $S_1$ and $S_2$
respectively should satisfy the values computed in
Lemma~\ref{lem:two-disjoint}, else the solution can be
improved. \eat{else the optimal solution
can be improved by redistributing the subspectrum/power
within $S_1$ and $S_2$.} Thus, we have:
\cbl
\begin{equation}
\frac{S_1}{S_2}=\frac{N_2P_1\hoo}{N_1P_2\htt} \label{eqn:fourth}
\end{equation}
\cb

\item
Below, we show how to derive the remaining two equations, which require
some tedious analysis.
\end{itemize}

\softpara{Remaining Two Equations (Eqns~\ref{eqn:fifth}-\ref{eqn:sixth}).}
Let us now consider a
small portion of the spectrum called $S$ --- taken partly from $S_1$
and $S_{12}$. In an optimal solution, redistribution of power within
$S$ should not lead to an improved total capacity. Without any loss of
generality, let us assume $S$ to be of size $(w+1)$, with $w>0$ in the
exclusive part ($S_1$) and $1$ in the shared part ($S_3$). See
Figure~\ref{fig:overlap}.
Thus, the total power used by the first user in $S$ is $w(c_1
+\sigma_1)+\sigma_1$. Let the {\em optimal} distribution of this total
power for user 1 within $S$ be in the ratio of $k:(1-k)$ ($0 \leq k
\leq 1$) between the exclusive and shared parts of $S$. Now, the total
capacity of both users in $S$ for the above power distribution is
given by:
\[ \begin{array}{ll}
C(k) = &  w\log (1+\frac{k(w(c_{1} +\sigma_{1} )+\sigma_{1})\hoo}{wN_{1} } ) + \\
 & \log (1+\frac{(1-k)(w(c_{1} +\sigma_{1} )+\sigma_{1} )\hoo}{\sigma_{2}\hto +N_{1} } )+ \\
   & \log (1+\frac{\sigma_{2}\htt}{\hot(1-k)(w(c_{1} +\sigma_{1} )+\sigma_{1} )+N_{2} })
\end{array} \]
Since $C(k)$ is connected and derivable for $0 \le k \le 1$, $C(k)$
can be optimal only at $k$ = 0, 1, or when $\frac{dC}{dk} = 0$.
Having $k$ = 0 or 1 will contradict our choice of $S$; thus,
$\frac{dC(k)}{dk}$ must be zero at optimal $C(k)$. Since we started
with an optimal SAPD solution, where the capacity $C(k)$ must also be
optimal, the value of $\frac{dC}{dk}$ must be zero for the
$k=\frac{w(c_{1} +\sigma_{1})}{w(c_{1} +\sigma_{1} )+\sigma_{1}}$
(based on the distribution of power in the original solution), and
this must be true for any $w$ in $(0,x]$ where $x$ is the size
of $S_1$ (the exclusive part of the spectrum).

{\em Analyzing $dC(k)/dk$.}  We computed $\frac{dC(k)}{dk}$ at
$k=\frac{w(c_{1}+\sigma_{1})}{w(c_{1} +
  \sigma_{1})+\sigma_{1}}$. After simplification, the numerator in the
resulting expression can be written as $w(\sigma_1+c_1)\Gamma_1+\sigma_1\Gamma_1$,
where
\[
\begin{array}{ll}
    \Gamma_1 =& {\htt}N_1^2\sigma_2+2{\htt\hoo}N_1\sigma_1\sigma_2+{\htt\hoo}c_1N_1\sigma_2\\
    & +\ {\htt}N_1\sigma_2^2-c_1N_2^2{\hoo}^2+c_1{\hoo\htt}\sigma_2^2\\
    & -\ c_1{\htt}N_2{\hoo}^2\sigma_2+2N_2{\hoo}\sigma_1\sigma_2+{\htt}N_2{\hoo}\sigma_2^2\\
    & +\ {\htt\hoo}^2\sigma_1^2\sigma_2-c_1{\hoo}^2\sigma_1^2+{\hoo}\sigma_1^2\sigma_2\\
    & +\ 2{\htt\hoo}\sigma_1\sigma_2^2+ N_2^2{\hoo}\sigma_2-2c_1N_2{\hoo}^2\sigma_1
\end{array}
\]
Since the numerator of $\frac{dC(k)}{dk}$ should be zero regardless of
$w$'s value in $(0,x]$, we must have that $\Gamma_1$ is
  zero. Similarly, for user 2, we must have $\Gamma_2 = 0$, where
  $\Gamma_2$ is similarly defined as $\Gamma_1$. Thus, we get the
  fifth and sixth equations as:
\begin{eqnarray}
\Gamma_1 &=& 0 \label{eqn:fifth} \\
\Gamma_2 &=& 0 \label{eqn:sixth}
\end{eqnarray}

\softpara{Eliminations of Variables.}  It is easy to verify that the
derived six equations are independent, and hence, are sufficient to
eliminate six (out of the total seven) variables as desired. However,
the order of elimination needs to be chosen carefully chosen to avoid
getting into a unsolvable polynomial of high degree. We choose
the following order of elimination. From Equation~\ref{eqn:first}, we
get:
\[
S_{12}=W-S_1-S_2
\]
Substituting the above in Equation~\ref{eqn:second} and~\ref{eqn:third},
and solving the resulting two equations for $S_1$ and $S_2$, we get
\begin{eqnarray*}
S_1&=&\frac{-W\sigma_1^2+P_2\sigma_1+P_1c_2-Wc_2\sigma_2}{c_1c_2-\sigma_1\sigma_2}\\
S_2&=&\frac{-W\sigma_2^2+P_1\sigma_2+P_2c_1-Wc_1\sigma_1}{c_1c_2-\sigma_1\sigma_2}
\end{eqnarray*}
We can now write Equation~\ref{eqn:fourth} as follows.
\[
\frac{\frac{-W\sigma_1^2+P_2\sigma_1+P_1c_2-Wc_2\sigma_2}{c_1c_2-\sigma_1\sigma_2}}{\frac{-W\sigma_2^2+P_1\sigma_2+P_2c_1-Wc_1\sigma_1}{c_1c_2-\sigma_1\sigma_2}}=\frac{N_2P_1\hoo}{N_1P_2\htt}
\]
In the above equation, we substitute $c_1$ and $c_2$ by the
expressions derived from Equations~\ref{eqn:fifth} and~\ref{eqn:sixth}
respectively.  Note that Equations~\ref{eqn:fifth} and~\ref{eqn:sixth}
are linear in $c_1$ and $c_2$ respectively, and hence, facilitating
the above substitutions. After the above substitutions and tedious
simplications, we actually get a fourth-degree equation in $\sigma_1$
(in terms of $\sigma_2$). Since four-degree equations have closed-form
solutions, we solve the resulting equation to express $\sigma_1$ in
terms of $\sigma_2$. The resulting expressions are extremely long and
tedious, and hence omitted here (see~\cite{matlab} for details).
The above allows us to express $B$ solely in terms of $\sigma_2$.
Thus, the single-variable equation $dB/d(\sigma_2)= 0$ can be solved
efficiently using well-known numerical methods, since $dB/d(\sigma_2)$
is connected and derivable in $\sigma_2$ with bounded derivatives, and
$\sigma_2$ has a bounded range (see Appendix~\ref{app:bound}). Finally, as $B$ is
continuous and bounded, we can then use the roots of $dB/d(\sigma_2)=
0$ to compute the optimal $B$.

\softpara{Note on Multiple Roots.}  Note that some of the intermediate
equations in the above described process may not be linear, and hence
may yield multiple roots. That only results in multiple expressions
for $B$ (in terms of $\sigma_2$), and hence, multiple possible sets
(but, at most 16 sets) of parameter values. We compute the total
system capacity $B$ for each of these set of values, and pick the one
that yields the largest value of $B$.

\section{\bf Conclusions}
\label{sec:conc}

In this paper, we have considered the spectrum management problem in
multiuser communication systems. We proved that in an optimal
solution, each user uses the maximum power. For the special case of
two users in flat channels, we solve the problem optimally.  Our
future work is focussed on generalization of our techniques
to communication systems with more than two users.

\eat{In the context
  of maximizing wireless network capacity, the SAPD problem addressed
  in the paper applies directly to wireless networks with one-hop
  traffic flows that are {\em saturated.} Using similar techniques as
  developed in this work, one of our future research directions is to
  design practical algorithms for multiple link networks with given
  traffic flows.}

\end{spacing}
\bibliographystyle{plain}
\bibliography{mobicom12}

\begin{thebibliography}{10}

\bibitem{matlab}
Matlab source codes for omitted mathematical details.
\newblock http://tinyurl.com/b6yhdal.

\bibitem{otherUCLA}
S.~R. Bhaskaran, Stephen~V. Hanly, Nasreen Badruddin, and Jamie~S. Evans.
\newblock Maximizing the sum rate in symmetric networks of interfering links.
\newblock {\em IEEE Trans.\ on Information Theory}, 2010.

\bibitem{cendrillon2006optimal}
R.~Cendrillon, W.~Yu, M.~Moonen, J.~Verlinden, and T.~Bostoen.
\newblock Optimal multiuser spectrum balancing for digital subscriber lines.
\newblock {\em Communications, IEEE Transactions on}, 54(5):922--933, 2006.

\bibitem{stanfordPower}
M.~Charafeddine and A.~Paulraj.
\newblock Maximum sum rates via analysis of 2-user interference channel
  achievable rates region.
\newblock In {\em IEEE CISS}, 2009.

\bibitem{binary-emk}
M.~Ebrahimi, M.~A. Maddah-ali, and A.~K. Khandani.
\newblock Power allocation and asymptotic achievable sum-rates in single-hop
  wireless networks.
\newblock In {\em IEEE CISS}, 2006.

\bibitem{ucb}
R.~H. Etkin, A.~Parekh, and D.~Tse.
\newblock Spectrum sharing for unlicensed bands.
\newblock {\em IEEE JSAC}, 25(3), 2007.

\bibitem{binary-ggok}
A.~Gjendemsj, D.~Gesbert, G.~E. Oien, and S.~G. Kiani.
\newblock Binary power control for sum rate maximization over multiple
  interfering links.
\newblock {\em IEEE Trans.\ Wireless Communications}, 7(8), 2008.

\bibitem{otherPower}
A.~Gjendemsjo, D.~Gesbert, G.E. Oien, and S.G. Kiani.
\newblock Optimal power allocation and scheduling for two-cell capacity
  maximization.
\newblock In {\em Intl.\ Symp.\ on Modeling and Optimization in Mobile, Ad Hoc
  and Wireless Networks}, 2006.

\bibitem{gummadi-hotnets-08}
R.~Gummadi, R.~Patra, H.~Balakrishnan, and E.~Brewer.
\newblock Interference avoidance and control.
\newblock In {\em ACM HotNets}, 2008.

\bibitem{matrix}
Himanshu Gupta and P~Sadayappan.
\newblock Communication efficient matrix multiplication on hypercubes.
\newblock In {\em Proceedings of the sixth annual ACM symposium on Parallel
  algorithms and architectures}, pages 320--329. ACM, 1994.

\bibitem{umn-2}
S.~Hayashi and Z.-Q. Luo.
\newblock Spectrum management for interference-limited multiuser communication
  systems.
\newblock {\em IEEE Trans. Inf. Theor.}, 2009.

\bibitem{umn-1}
Z.-Q. Luo and S.~Zhang.
\newblock Dynamic spectrum management: Complexity and duality.
\newblock {\em IEEE J.\ of Selected Topics in Signal Processing}, 2008.

\bibitem{umn-11}
Z.-Q. Luo and S.~Zhang.
\newblock Duality gap estimation and polynomial time approximation for optimal
  spectrum management.
\newblock {\em ACM Trans. Sig. Proc.}, 2009.

\bibitem{hong}
H.~Shen, Hang Zhou, R.A. Berry, and M.L. Honig.
\newblock Optimal spectrum allocation in gaussian interference networks.
\newblock In {\em Asilomar Conf. on Signals, Systems and Computers}, 2008.

\bibitem{yu2002distributed}
W.~Yu, G.~Ginis, and J.M. Cioffi.
\newblock Distributed multiuser power control for digital subscriber lines.
\newblock {\em IEEE JSAC}, 2002.

\bibitem{yu2006dual}
W.~Yu and R.~Lui.
\newblock Dual methods for nonconvex spectrum optimization of multicarrier
  systems.
\newblock {\em IEEE Trans.\ on Comm.}, 2006.

\bibitem{UCLApaper}
Y.~Zhao and G.J. Pottie.
\newblock Optimal spectrum management in multiuser interference channels.
\newblock In {\em ISIT}, 2009.

\bibitem{join}
Xianjin Zhu, Himanshu Gupta, and Bin Tang.
\newblock Join of multiple data streams in sensor networks.
\newblock {\em IEEE Transactions on knowledge and Data Engineering},
  21(12):1722--1736, 2009.

\bibitem{msn}
Xianjin Zhu, Bin Tang, and Himanshu Gupta.
\newblock Delay efficient data gathering in sensor networks.
\newblock In {\em International Conference on Mobile Ad-Hoc and Sensor
  Networks}, pages 380--389. Springer, 2005.

\end{thebibliography}
\appendices
\section{\bf Proof of Theorem~\ref{thm:multi-power-fair}}
\label{app:fair}

\para{Proof of Theorem~\ref{thm:multi-power-fair}.}
We make the same changes as suggested in Theorem~\ref{thm:multi-power}'s proof.
The suggested changes will result in the objective value changing from
$$(\Pi_i w_iC_i)\ {\rm to}\ w_k(C_k + \ckp - \bigtriangledown_k) \Pi_{i \neq k} w_k(C_i - \bigtriangledown_i),$$
where $w_i$ and $C_i$ are the weights and total capacity of user $i$. 
Note that $(C_k - \bigtriangledown_k) \geq 0.$  Let $\eta'$ be the ratio of
the above objective values (new to old value). Below, we show that
there exists an $\epsilon$ that makes $\eta' > 1$. This would imply
that the given optimal solution is suboptimal (a contradiction),
and thus, proving the theorem. 

Now, using Eqn~\ref{eqn:dec} and~\ref{eqn:inc}, we get:
\begin{eqnarray*}
\eta' &\triangleq&  \left(\prod_{i \neq k}{\frac{C_i-\bigtriangledown_i}{C_i}}\right){\frac{C_k+(\ckp- \bigtriangledown_k)}{C_k}} \\
     &\geq& \left(\prod_{i \neq k}{\frac{C_{min}-\bigtriangledown_i}{C_{min}}}\right){\frac{C_{max}+(\ckp- \bigtriangledown_k)}{C_{max}}} \\
     &\geq& \left(\frac{C_{min}-\bigtriangledown_{max}}{C_{min}}\right)^{n-1}{\frac{C_{max}+(\ckp- \bigtriangledown_k)}{C_{max}}} \\
     &\geq& (1-a_1\epsilon)^{n-1}(1+a_2\epsilon \log(1+\frac{a_3}{\epsilon})-a_4\epsilon)
\end{eqnarray*}
where $a_1,a_2,a_3,a_4$ are appropriate {\em positive} constants (independent of $\epsilon$) and $\bigtriangledown_{max}$ is the
expression in Equation~\ref{eqn:dec}. Let $\eta$ denote the last expression above.
We can now state the following:
\[
\begin{array}[t]{ll}
    (i)  & \lim_{\epsilon \to 0} \eta=1. \\
    (ii) & \begin{array}[t]{rl} \frac{d\eta}{d\epsilon}=& (1-a_1{\epsilon})^{n-1}\times\\
    &\Bigg( \frac{-a_1(n-1)\left( 1+a_2\epsilon\log(1+a_3/\epsilon)-a_4\epsilon \right)}{1-a_1{\epsilon}}+ \\
    &a_2\log(1+a_3/\epsilon) - \frac{a_2\epsilon}{(1+a_3/\epsilon)\epsilon^2} - a_4 \Bigg) \\
    =&(1-a_1{\epsilon})^{n-1}.\xi \\
    \end{array}
\end{array}
\]
Also, one can easily verify that $\lim_{\epsilon \to 0^+}
\xi= +\infty$ and $(1-a_1{\epsilon})^{n-1}$ is always positive. Thus,
$\frac{d\eta}{d\epsilon}$ is positive when $\epsilon \to 0^+$, which
implies (from (i) above) that there exists an $\epsilon > 0$ such
that $\eta > 1$ and thus $\eta' > 1$.
\QD

\section{\bf Proof of Lemma~\ref{lem:rect}}
\label{app:rect}
\para{Proof of Lemma~\ref{lem:rect}.}
\cbl Instead of directly proving Lemma~\ref{lem:rect}, 
we prove the following lemma.\cb

\begin{lemma}
\label{lem:rect-aux}
Consider two users 1 and 2, and an SAPD solution (not necessarily
optimal) $\{p_1(x), p_2(x)\}$ where each user uses the entire
available spectrum $[0,W]$.
We claim that there always exists an SAPD solution $\{p'_1(x),
p'_2(x)\}$ with equal or higher total capacity such that either (i)
both the PSD functions $p'_i(x)$ are constant in $[0,W]$, or (ii) one
of the users does not use the entire spectrum $[0,W]$.
\end{lemma}

\cbl
Lemma~\ref{lem:rect} can be easily inferred from
Lemma~\ref{lem:rect-aux} by using contradiction. Lets consider an SAPD
problem instance for two users, which has no optimal solution wherein
the PSDs of the two users is constant in the shared part of the
spectrum. From the set of optimal solutions, lets pick the one with
minimum size of the shared spectrum. According to
lemma~\ref{lem:rect-aux}, we can find another solution with equal or
higher capacity in which either the size of the shared spectrum is
reduced or the users use constant PSD's in the shared spectrum. In
either case, we get a contradiction. We now present the proof of
Lemma~\ref{lem:rect-aux}.
\cb

\para{Proof of Lemma~\ref{lem:rect-aux}.}
We start with defining a couple of notations.

\softpara{$k$-rectangular SAPD Solution}.
An SAPD solution $\{p_1(x), p_2(x)\}$ is considered to be
$k$-rectangular if there exists frequency values $w_i$, such that
$0=w_0 < w_1 < w_2 < \ldots < w_{k-1} < w_k = W$ such that for each
$j$ ($1 \leq j \leq k$) and $x$ ($w_{j-1} \leq x < w_j$), we have
$p_1(x)=c_{1j}$ and $p_2(x)=c_{2j}$ for some constants $c_{1j}$ and
$c_{2j}$.

\para{2-rectangular SAPD Solution.}
First, we prove the lemma for the special case when the given SAPD
solution $\{p_1(x), p_2(x)\}$ is $2$-rectangular. Without loss of
generality, let us assume that the given SAPD solution is the {\em
optimal} $2$-rectangular SAPD solution, under the given total powers
(viz., $\int_{0}^{W}p_1(x)dx$ and $\int_{0}^{W}p_2(x)dx$
respectively). Now, we can write the given optimal $2$-rectangular SAPD
solution as follows.
\begin{itemize}
\item
For $0 \leq x < w$, $p_1(x) = \sigma_1$, $p_2(x) = \sigma_2$.
\item
For $w \leq x < W$, $p_1(x) = \sigma_1 + \Delta_1$, $p_2(x) = \sigma_2 + \Delta_2$.
\end{itemize}
Above, $\sigma_i > 0$, $\Delta_i + \sigma_i > 0$, for each $i$.
Let $\Psi_1$ and $\Psi_2$ be the aggregate (sum over two links)
capacity per unit-bandwidth in the two sub-spectrums $[0,w]$ and
$(w,W]$ respectively. Without loss of generality, let us assume
$\Psi_1 \leq \Psi_2$. We consider the following four cases.

\softpara{$\Psi_{1} = \Psi_{2} = \Psi$ and $\Delta_{1}\Delta_{2} = 0$.}
In this case, the given solution can be easily converted to a
1-rectangular solution of equal or higher capacity.

\softpara{$\Psi_{1} = \Psi_{2} = \Psi$ and $\Delta_{1}\Delta_{2} > 0$.}
Without loss of generality, we assume $\Delta_{2} \geq \Delta_{1} >
0$.\footnote{If both are negative, then we can reverse the role of the
two sub-spectrums.} Note that, in either sub-spectrum, if we
``scale-up'' the PSD value of each link, then the aggregate capacity
(per unit-bandwidth) would increase.
Thus, for any $a>1$, the PSD value of $a.\sigma_{1}$ and
$a.\sigma_{2}$ would result in a higher aggregate capacity than
$\Psi_1$ (= $\Psi_{2})$. Now, since $\Delta_i > 0$, there exists $a>1$
such that $a.\sigma_i < \sigma_i+\Delta_i$ for each $i$. For such an
$a$, changing the PSD value in the second sub-spectrum from
$\sigma_i+\Delta_i$ to $a\sigma_i$ results in an increase in the
aggregate capacity (with lower total power). Thus, the given solution
is not an optimal 2-rectangular solution. QED.

\softpara{$\Psi_{1} = \Psi_{2}= \Psi$ and $\Delta_{1}\Delta_{2} < 0$.}
Without loss of generality, we can assume $\Delta_{1} > 0$ and
$\Delta_{2} < 0$. Now, if $W>2w$, let $[g_1, g_2]=[0,2w]$ otherwise
let $[g_1,g_2]=[W-2w,W]$.
Let $X(b)$ and $Y(b)$ be such that $\log X(b)$ and $\log Y(b)$ are the
capacities per unit-bandwidth of the first and second links when they
use a constant PSD value of $\sigma{1}+b\Delta{1}$ and
$\sigma{2}+b\Delta{2}$ respectively; here, $b \in
[-\frac{\sigma{1}}{\Delta_{1}}, -\frac{\sigma{2}}{\Delta{2}}]
\supseteq [0,1]$.
Below, we show how to choose appropriate $b$ values to create a better
2-rectangular solution, or an equal-capacity solution wherein one
of the links does not use the entire spectrum.

\begin{figure}[thbp]
\label{fig::diag2}
\vspace*{-1.3in}
\hspace{-.4in}\includegraphics[width=5in]{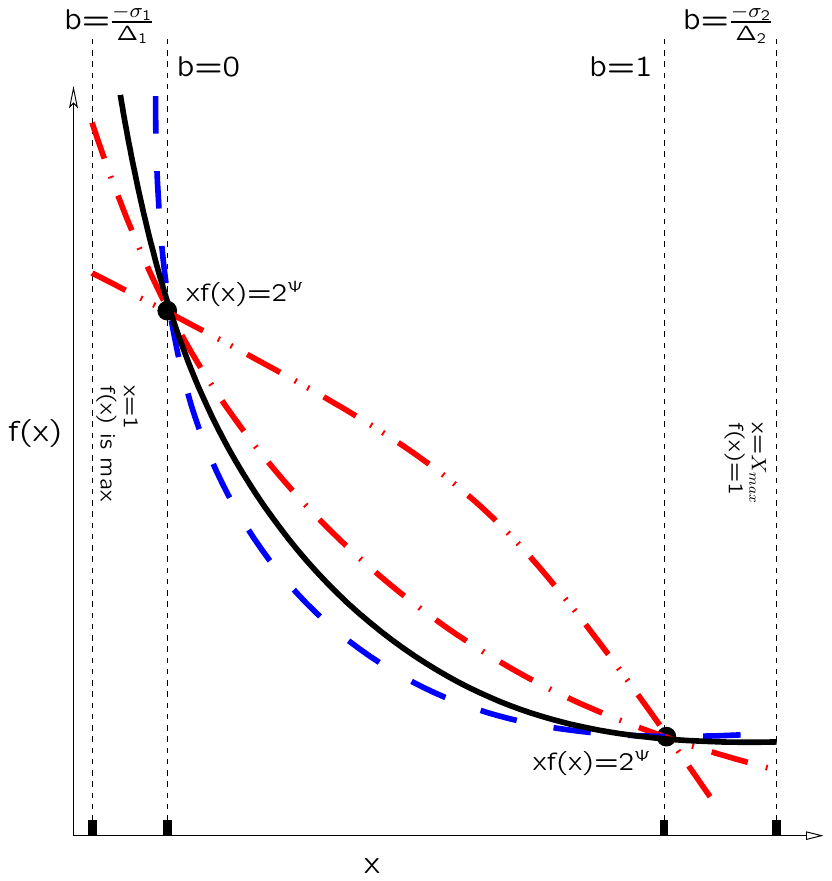}
\vspace*{-2in}
\caption{\footnotesize Red and blue (dotted) curves are
the possible shapes of $f(x)$; here, the black (solid) curve is $y =
2^{\Psi}/x$.}
\label{fig:diag2}
\vspace*{-0.1in}
\end{figure}

Let $X_{max}$ be the maximum value of $X(b)$ over the above range of $b$.
Since the above function $X(b)$ is reversible, we can define the
function $f = Y(X^{-1}): [0,X_{max}] \mapsto \mathbb{R}_{\geq 0}$ such
that $f(x)$ gives the capacity-per-bandwidth of the second link when
the capacity/bandwidth of the first link is $x$ {\em due to} constant
PSD values of $\sigma{1}+b\Delta{1}$ and $\sigma{2}+b\Delta{2}$
respectively for some $b$; note that, $b$ is unique for a given
$x$. We can show (we omit the details here) that the second-derivative
of the function ($d(d f(x)/dx)/dx$) cannot be zero in $[0,X_{max}]$.
Thus, the function $f(x)$ has no inflection point in the range
$[0,X_{max}]$, and hence, we can plot the various possibilities for
the $f(x)$ relative to $y=2^{\Psi}/x$ as shown in
Figure~\ref{fig:diag2}.  Note that $f(x)$ is maximum at $x=1$, and is
1 at $X_{max}$, and intersects the $y=2^{\Psi}/x$ plot at two $x$
values corresponding to $b=0$ and $b=1$ (since $\Psi_{1} =
\Psi_{2}= \Psi$). Moreover, since $X(b)$ is monotonically increasing
in $b$, we get the values/ranges of $b$ as depicted in the
figure. Now, for each of the four possibilities of $f(x)$ depicted in
the Figure~\ref{fig:diag2}, we can prove the lemma as follows.
\begin{itemize}
\item
If $f(x)$ is one of the two red plots, then we pick $b=1/2$. For
$b=1/2$, we get $X(b)Y(b) > 2^\Psi$ and hence $\log X(b) + \log Y(b) >
\Psi$. Now, if we can choose constant PSD values of
$\sigma{1}+b\Delta{1}$ and $\sigma{2}+b\Delta{2}$ for the two links
respectively in $[g_1, g_2]$, we get a 2-rectangular solution in
$[0,W]$ of higher total capacity within the given power
constraint. QED.

\item
If $f(x)$ is the blue or the black plot, then we choose two values of
$b$, viz., $b_l$ and $b_r$, so as to use PSD values of $\sigma_i + b_l
\Delta_i$ in $[g_1,w]$ and $\sigma_i + b_r \Delta_i$ in $[w,g_2]$
for each link $i$. For our purposes, we need to choose $b_l$ and $b_r$
such that they satisfy the following three conditions: (i)
$-\sigma_1/\Delta_1 \leq b_l \leq 0$, and $1 \leq b_r \leq
-\sigma_2/\Delta_2$ (to ensure that $b$ is in the valid range and the
capacity/bandwidth is at least $\Psi$ in each sub-spectrum), and (ii)
$b_l + b_r = 1$ (to ensure that the total power used is at most the
total power in the original solution, for each link), and (iii)
$\sigma_i + b_l \Delta_i$ or $\sigma_i + b_r \Delta_i$ is zero for
some $i$ (so that one of the links uses zero power in one of the
sub-spectrums). To satisfy the above three conditions, we choose the
pair $(b_l,b_r)$ as $(-\sigma_1/\Delta_1, 1 +
\sigma_1/\Delta_1)$ if $1 + \sigma_1/\Delta_1 < - \sigma_2/\Delta_2$,
or $(1+\sigma_2/\Delta_2, -\sigma_2/\Delta_2)$ otherwise.
The above yields an SAPD solution of higher capacity wherein one of
the links doesn't use the entire spectrum. QED.
\end{itemize}

\softpara{$\Psi_{1} < \Psi_{2}$.}
In this case, we consider sub-spectrums $[g_1,w]$ and $[w,g_2]$ for
some appropriate $g_1$ and $g_2$ (determined later), and increase the
aggregate capacity within these sub-spectrums by appropriate
redistribution of power.

Let $r=(g_2-w)/(w-g_1)$, the ratio of the two sub-subspectrums, and
$\tau = (g_2-g_1)$.
Let $P'_1$ and $P'_2$ be the total power used by link $1$ and $2$ in
$[g_1,g_2]$, i.e., $P'_i={\tau}(\sigma_{i}+ \Delta_ir/(r+1))$.  Let
$\Phi(r)$ be the aggregate capacity per bandwidth in $[g_1,g_2]$ when
the PSD values are $P'_1/\tau$ and $P'_2/\tau$ respectively for the
two links. We now show that a ``large-enough'' $r$ will ensure that
$(1 + r)\Psi(r) > \Psi_1 + r\Psi_2$, which will imply that in
$[g_1,g_2]$ the 1-rectangular solution yields a higher total capacity
than the given solution.
Observe the following: (i) $\lim_{r \rightarrow 0^+} {\Phi(r) =
\Psi_{1}}$, (ii) $\lim_{r \rightarrow \infty}{\Phi(r) =
\Psi_{2}}$, and (iii) $\phi(r)$ is connected. Since $\lim_{r
\rightarrow \infty^+}{(1+r) \Phi(r)=(1+r)\Psi_{2}>\Psi_{1}+r
\Psi_{2}}$, there exists a large-enough $r$ for which $(1
+ r)\Psi(r) > \Psi_{1}+r\Psi_{2}$.
Once we find the appropriate $r$, we can determine $g_1$ and $g_2$ as
follows: If $(r+1)w<W$, then pick $[g_1,g_2] = [0,(r+1)w]$, else pick
$[g_1,g_2] = [w - (W-w)/r, W]$.
Then, in $[g_1,g_2]$, we use power-signals of $\sigma_i+\Delta_i
r/(r+1)$ for link $i$, yielding a 2-rectangular solution with a
higher-capacity than the given solution. QED.

\para{$k$-rectangular Solution.}
This can be easily proven by induction on $k$, using the above result
on $k=2$ as the base case.

\para{Arbitrary SAPD Solution.}
Let $p_1(x)$ and $p_2(x)$ be the power-distribution functions for the
given solution, and let $P^*_i=\int_{0}^{W}p_i(x)dx$ be the total powers
used by the links. Assume that there is no solution of equal or higher
capacity, in which one of the link doesn't use the full spectrum.
Let us construct an $n$-rectangular solution that ``approximates'' the
given solution as follows: First, we divide the spectrum $[0,W]$ into
$n$ equi-sized sub-spectrums, and then, within each sub-spectrum we
use a constant PSD value of {\em minimum} $p_i(x)$ in that
sub-spectrum.  Note that the total power used by the link $i$ in the
above $n$-rectangular solution is atmost $P^*_i$.  Let $F_n$ be the
total capacity of the above $n$-rectangular solution, and let $R$ be
the total capacity of the 1-rectangular solution that uses a constant
PSD of $P_i/W$ for each link. Since the lemma holds for
$k$-rectangular solutions, we get that $F_n \leq R$ for any $n$.  Now,
if $C$ is the total capacity of the given solution, then by definition
$C = \lim_{n \rightarrow +\infty} F_n$. Thus, we get $C \leq R$,
which completes the proof.
\QD

\eat{
We prove this by induction on $k$. Above, we have proven the base case
for $k=2$. Let us assume that the lemma is true for any
$c$-rectangular solution for $1<c<k$. In a $k$-rectangular solution
with $0=w_0<w_1<w_2<\ldots<w_k=W$, choose the sub-spectrum $[0,w_2]$
and apply the lemma for the base case of $k=2$. This either yields a
equal or higher-capacity solution that is either $(k-1)$-rectangular
or wherein one of the links doesn't use the full spectrum. In the
former case, the lemma holds by the inductive hypothesis, while the
latter case proves the lemma directly.

\begin{itemize}
\item
$\log X(b) + \log Y(b) > \Psi$ for $b=1/2$. In this case, we can use
$\sigma_i + b \Delta_i$ PSD values in $[g_1,g_2]$ to yield a better
2-rectangular solution within the given power constraint, and
thus, proving the theorem.

\item
Two values $b_l$ and $b_r$, such that (a) $b_l + b_r = 1$, (b)
$\sigma_i + b_l \Delta_i$ and $\sigma_i + b_r \Delta_i$ are positive
for $i = 1,2$, (c) $\log X(b_l) + \log Y(b_l) > \Psi$ and $\log X(b_r)
+ \log Y(b_r) > \Psi$, and (d) $\sigma_i + b_j \Delta_i$ = 0 for some
$i \in  {1,2}$ and $b_j \in \{b_l, b_r\}$.
\end{itemize}
}

\section{\bf Proofs of Lemma~\ref{lem:two-disjoint} and~\ref{lem:diff-inter}}
\label{app:disjoint}

\para{Proof of Lemma~\ref{lem:two-disjoint}.}
First, it is easy to see that the union of the disjoint spectrums must
be the entire available spectrum.  Let the links use disjoint
spectrums of size $yW$ and $(1-y)W$ where $0 \leq y \leq 1$.  Since
both links should use maximum power for maximum capacity, we can
compute the total capacity as follows.
\fontsize{8.5pt}{9pt}
\[C=yW\log (1+\frac{P_{1}\hoo}{yWN_{1} } )+W(1-y)\log (1+\frac{P_{2}\htt}{(1-y)WN_{2}})\]
\normalsize
We can find the optimal value of $y$ by solving for $dC/dy=0$. We have:
\[ \begin{array}{ll}
\frac{dC(y)}{dy} = & W \big( \log (1+\frac{P_{1}\hoo}{yWN_{1} } )-\log (1+\frac{P_{2}\htt}{(1-y)WN_{2} } )- \\
&\frac{P_{1}\hoo}{yWN_{1} (1+\frac{P_{1}\hoo}{yWN_{1} } )} +\frac{P_{2}\htt}{(1-y)WN_{2} (1+\frac{P_{2}\htt}{(1-y)WN_{2} } )} \big)
\end{array}\]
The root of the equation $dC/dy=0$ is:
\[y=\frac{N_{2} P_{1}\hoo}{N_{1} P_{2}\htt +N_{2} P_{1}\hoo}. \]

\cbl Hence, the PSD's of link $1$ and $2$ are $\frac{N_{1} P_{2}\htt +N_{2} P_{1}\hoo}{WN_{2}\hoo}$ and
$\frac{N_{1} P_{2}\htt +N_{2} P_{1}\hoo}{WN_{1}\htt}$ respectively \cb and the optimal value of $C$ is:
\[C=W\log (1+\frac{P_{1}\hoo}{WN_{1}} +\frac{P_{2}\htt}{WN_{2}})\]
\QD

\para{Lemma~\ref{lem:diff-inter}.}
\begin{lem-prf}
\label{lem:diff-inter}
Consider a communication system with a single user 1, and an available
spectrum $[0,W]$. Let the interference (from other users) in the
sub-spectrums $[0,w]$ and $(w,W]$ be constant and equal to $I$ and
$I'$ respectively. If $I > I'$, then to achieve maximum capacity for
user 1, its PSD value in $[0,w]$ should be lower than in $(w,W]$.
\end{lem-prf}
\begin{proof}
It is easy to see that for optimal capacity: (i) the PSD should be
constant in each of the sub-spectrums, and (ii) the link should use
maximum power.  Now, if we divide the total power of $P_1$ into the
two sub-spectrums in the ratio of $k:(1-k)$, for some $0 \leq k \leq
1$, we get link capacity as:
\fontsize{8.5pt}{9pt}
\[C(k)=w\log (1+\frac{kP_{1}\hoo}{w(I + N_1) } )+(W-w)\log (1+\frac{(1-k)P_{1}\hoo}{(W-w)(I' + N_1) } )\]
\normalsize
By solving $dC/dk=0$, we get
$k = \frac{w}{W}+ \frac{w}{WP_1\hoo}(I' - I)(W-w)$
which give us the PSD values of
$\frac{1}{W}(P_1 + (W-w)(I' -I)/\hoo)$ and
$\frac{1}{W}(P_1 +w(I -I')/\hoo)$ in
the two sub-spectrums. \blue{This proves the lemma, since the first
PSD value is always greater than the second PSD value.}
\end{proof}

\section{\bf Cases for $S_1$ or $S_2$ = 0.}  
\label{app:cases}

\softpara{Case where $S_1$ or $S_2$ is of Zero Size.}  Let $S_1=0$ and
$S_2 > 0$. In this case, Equations~\ref{eqn:fourth}
and~\ref{eqn:fifth} are not valid. At the same time, the variables
$S_1$ and $c_1$ are eliminated from the system, and hence, we have two
fewer equations and variables which only simplifies the problem. We
can use the exact same order of elimination and technique to yield an
optimal solution for this case. This case of $S_2=0$ and $S_1 > 0$ is
similarly handled, and the case of $S_2=0$ and $S_1 = 1$ is already
handled by Lemma~\ref{lem:full-overlap}.

\section{\bf Upper Bound of $\sigma_2$}
\label{app:bound}

\softpara{Upper bound of $\sigma_2$.}
Here, we show that there exists an upper bound for $\sigma_2$. Since the PSD's used by users
1 and 2 in $S_1$ and $S_2$ is $(c_1+\sigma_1)S_1$ and $(c_2+\sigma_2)S_2$ respectively, we have
the following (by applying Lemma~\ref{lem:two-disjoint}, and using the PSD values computed
therein):
\begin{align*}
&c_1+\sigma_1=\frac{N_{2}(c_1+\sigma_1)S_1\hoo +N_{1}(c_2+\sigma_2)S_2\htt}{(S_1+S_2)N_{2}\hoo}, \\
&c_2+\sigma_2=\frac{N_{2}(c_1+\sigma_1)S_1\hoo +N_{1}(c_2+\sigma_2)S_2\htt}{(S_1+S_2)N_{1}\htt},
\end{align*}
and $c_2+\sigma_2=(c_1+\sigma_1)\frac{N_{2}\hoo}{N_{1}\htt}$. Let
$\varsigma=\frac{N_{2}\hoo}{N_{1}\htt}$ and $\gamma=\max(\varsigma,1)$.
Let $P=P_1+P_2$, and recall that $c_1,c_2,\sigma_1$, and $\sigma_2$ are positive numbers.
Thus, we have:
\begin{eqnarray*}
P &=& (c_1+\sigma_1)S_1+(c_2+\sigma_2)S_2+(\sigma_1+\sigma_2)S_{12} \\
P &>& \frac{1}{\varsigma}(c_2+\sigma_2)S_1+\sigma_2S_2+\sigma_2S_{12} \\
\gamma P &>& (\gamma/\varsigma)\sigma_2S_1 + \gamma\sigma_2(S_2+S_{12}) \\
\gamma P &>& \sigma_2 (S_1 + S_2 + S_{12})  \ \ \ (\rm{as}\ \gamma \geq 1, \varsigma) \\
\gamma P/W &>& \sigma_2
\end{eqnarray*}
Thus, $\gamma P/W$ is an upper bound on $\sigma_2$, where $\gamma = \max(1, \frac{N_{2}\hoo}{N_{1}\htt})$.

\end{document}